\documentclass[reprint,showpacs,preprintnumbers,amsmath,aps,pre]{revtex4-1}
\usepackage{graphicx}
\usepackage{color}

\begin{document}

\preprint{Commun. Theor. Phys. \textbf{59}, 175-178 (2013)}

\title{Bounds of efficiency at maximum power for normal-, sub- and super-dissipative Carnot-like heat engines}

\author{Yang Wang}
\affiliation{Department of Physics, Beijing Normal University, Beijing 100875, China}
\author{Z. C. Tu}\email{Corresponding author: tuzc@bnu.edu.cn}
\affiliation{Department of Physics, Beijing Normal University, Beijing 100875, China}

\begin{abstract}
The Carnot-like  heat engines are classified into three types (normal-, sub- and super-dissipative) according to relations between the minimum irreversible entropy production in the ``isothermal" processes and the time for completing those processes. The efficiencies at maximum power of normal-, sub- and super-dissipative Carnot-like heat engines are proved to be bounded between $\eta_C/2$ and $\eta_C/(2-\eta_C)$, $\eta_C /2$ and $\eta_C$, $0$ and $\eta_C/(2-\eta_C)$, respectively. These bounds are also shared by linear, sub- and super-linear irreversible Carnot-like engines [Tu and Wang, Europhys. Lett. \textbf{98}, 40001 (2012)] although the dissipative engines and the irreversible ones are inequivalent to each other.
\end{abstract}
\maketitle
\setcounter{page}{175}
\section{\label{sec:level1} Introduction}

Inspired by the pioneer achievements make by Yvon~\cite{Yvon}, Novilov~\cite{Novikov}, Chambadal~\cite{Chambadal},  Curzon and Ahlborn~\cite{Curzon1975}, the issue of efficiency at the maximum power (EMP) has drawn much attention from scientists with the desire to design engines that can provide sufficient power with higher efficiency.
The emerging theoretical advances in  this field \cite{Salamon80,Salamon80,Chen1989,ChenJC94,Bejan96,ChenL99,vdbrk2005,dcisbj2007,Schmiedl2008,Tu2008,Chenlin2009,
Esposito2009a,Esposito2009,Esposito2010,EspositoPRE10,GaveauPRL10,Izumida,wangx10,Velasco10,Abe2011,chenl2011,Izumida2012,
 Schmiedl07,TuW201108,WangHe,WangHe41148,Apertet} improve our understanding to the issue of EMP for heat engines and irreversible
thermodynamics. Recently, Esposito \textit{et al.} investigated the Carnot-like engines working in the low-dissipation region \cite{Esposito2010} and obtained the lower bound $\eta_-\equiv \eta_C/2$ and the  upper bound $\eta_+ \equiv \eta_C/(2-\eta_C)$ of EMP for this kind of engines, where $\eta_C$ is the Carnot efficiency. In addition, Gaveau and his coworkers proposed a concept of sustainable efficiency and proved that it has the upper bound ${1}/{2}$, based on which they also obtained the upper bound $\eta_+={\eta_C}/(2-\eta_C)$ for the efficiency of Carnot-like engines at maximum power \cite{GaveauPRL10}. Seifert argued that the upper bound ${1}/{2}$ for the sustainable efficiency holds only in the linear nonequilibrium region \cite{Seifert11}. Esposito \emph{et al.} \cite{Esposito201201} provided a nice example of single level quantum dot where the upper bound can vary from $1/2$ to $1$ with increasing the thermodynamic force, which to some extent supports Seifert's argument. Therefore the upper bound $\eta_C/(2-\eta_C)$ might not hold for EMP of Carnot-like heat engines arbitrarily far from equilibrium. In recent work, we classified irreversible Carnot-like heat engines into three types (linear, sublinear and superlinear), and derived the corresponding EMP to be bounded
between $\eta_C/2$ and $\eta_C/(2-\eta_C)$, $\eta_C /2$ and $\eta_C$, $0$ and $\eta_C/(2-\eta_C)$, respectively \cite{TuW201110}. Particularly, we found that the EMP of sublinear irreversible heat engines could reach the Carnot efficiency \cite{TuW201110}.

It is necessary to introduce the concept of effective temperature of working substance for most of models mentioned above. However, the definition of effective temperature is debatable in some cases. Esposito \textit{et al.} started from the time-dependent behavior of the irreversible entropy production in the ``isothermal"  (The quote marks on the word ``isothermal" merely indicate the working substance in contact with a reservoir at a constant temperature) process without introducing the effective temperature of working substance \cite{Esposito2010}, which inspired us that the upper and lower bounds of EMP might be straightly derived from the relation between the irreversible entropy production in the  ``isothermal" process and the time for completing that process.
Here, we classify the Carnot-like engines into three types (normal-, sub- and super-dissipative) according to the relations  between the minimum irreversible entropy production in the finite-time ``isothermal" processes and the time for completing those processes. The EMPs of normal-, sub- and super-dissipative Carnot-like heat engines are proved to be bounded between $\eta_C/2$ and $\eta_C/(2-\eta_C)$, $\eta_C /2$ and $\eta_C$, $0$ and $\eta_C/(2-\eta_C)$, respectively.

\section{theoretical model }
We consider a heat engine perform the following Carnot-like cycle which consists of four processes.

\textit{``Isothermal" expansion.}
The working substance of heat engine is in contact with a hot reservoir at temperature $T_1$ and the constraint on the system is loosened according to some external controlled parameter $\lambda_1(\tau)$ during the time interval $0<\tau<t_1$ where $\tau$ is the time variable. A certain amount of heat $Q_1$ is absorbed from the hot reservoir. Then the variation of entropy of working substance can be expressed as
\begin{equation}\label{Deltas_1}
    \Delta S_1= Q_1 / T_1+ \Delta S_1^{ir},
\end{equation}
where $\Delta S_1^{ir}\geq 0$ is the irreversible entropy production in this finite time process. Because $\Delta S_1>0$ is a state function, its value can be calculated from the variation of entropy in the real isothermal expansion process constituting an ideal Carnot cycle.

\textit{Adiabatic expansion.}
The adiabatic expansion is idealized as the working substance suddenly decouples from the hot reservoir and then comes into contact with the cold reservoir instantly at time $\tau=t_1$. During this transition, the constraint on the system is loosened further. There is no heat exchange and entropy production in this process.

\textit{``Isothermal" compression.}
The working substance is in contact with a cold reservoir at temperature $T_3$  and the constraint on the system is enhanced according to the external controlled parameter $\lambda_3(\tau)$ during the time interval $t_1<\tau<t_1+t_3$. A certain amount of heat $Q_3$ is released to the cold reservoir.  The variation of entropy of working substance in this process can be expressed as
\begin{equation}\label{Deltas_3}
    \Delta S_3=- Q_3/T_3+\Delta S_3^{ir},
\end{equation}
where $\Delta S_3^{ir}\geq 0$ is the irreversible entropy production. Because $\Delta S_3<0$ is also a state function, its value can be calculated from the variation of entropy in the real isothermal  compression process constituting the ideal Carnot cycle.

\textit{Adiabatic compression.}
Similar to the adiabatic expansion, the working substance suddenly decouples from the cold reservoir and then comes into contact with the hot reservoir instantaneously at time $\tau=t_1+t_3$. There is no heat exchange and entropy production in this process.

Having undergone this Carnot-like cycle, the system comes back to its initial state again. Thus there are no net energy change and variation of entropy in the whole cycle. Then we have $\Delta S_3=-\Delta S_1$ and
and the net work output $W=Q_1-Q_3$.
The power can be expressed as
\begin{equation}
 P=\frac{Q_1-Q_3}{t_{tot}}=\frac{(T_1-T_3)\Delta S_1-T_1\Delta S_1^{ir}-T_3\Delta S_3^{ir} }{t_1+t_3}.
\label{Pfull}
\end{equation}
According to the equation above, maximizing the power means minimizing the irreversible entropy production with respect to the protocols $\lambda_1(\tau)$ and $\lambda_1(\tau)$ for given time intervals $t_1$ and $t_3$ first, then Eq.~\eqref{Pfull} can be transformed into
\begin{equation}\label{pmodified}
    P=\frac{(T_1-T_3)\Delta S_1-T_1L_1-T_3L_3}{t_1+t_3},
\end{equation}
where $L_1$ and $L_3$ represent $\min\{\Delta S_1^{ir}\}$ and $\min\{\Delta S_1^{ir}\}$ for given time intervals $t_1$ and $t_3$, respectively. In fact, they reflect to what extent the engines depart from equilibrium for given time intervals $t_1$ and $t_3$.

\section{classification of dissipative heat engines}
Intuitionally, the irreversible entropy production decreases as the increase of time for completing the ``isothermal" processes, thus $L_1$ and $L_3$ can be expressed as monotonely increasing function with respect to $1/t_1$ and $1/t_3$, respectively. If we introduce a transformation of variables $x_i=1/t_i$ ($i=1,3$), the minimum irreversible entropy production in each ``isothermal" process can be expressed as $L_i = L_i(x_i)$ ($i=1,3$). This transformation has also been extended to investigate the optimizations of low-dissipation refrigerators~\cite{wangturefrig} and irreversible Carnot-like engines with internally dissipative friction~\cite{wangheinter}.
We can defined three kinds of typical characteristics according to the behavior of $L_i=L_i(x_i)$, which is similar to the analysis in our previous work \cite{TuW201110}. The first one is called normal dissipative type which is represented by the straight line in Fig.~1. This type of Carnot-like engines is also called the low-dissipation engines by Esposito \textit{et al.} \cite{Esposito2010}.  The second one is called sub-dissipative type which is represented by the convex curve in Fig.1. The third one is called super-dissipative type which is represented by the concave curve in Fig.1. The behavior of three kinds of characteristics can be mathematically expressed as
\begin{equation}\label{threetypes}
    \begin{cases}
     xL'  = L, & \text{normal-dissipative} \\
     xL'  > L, & \text {sub-dissipative}\\
     xL'  < L, & \text {super-dissipative}
    \end  {cases}
\end{equation}
where $L$, $x$, and $L'$ represent $L_1$(or $L_3$), $x_1$(or $x_3$) and $\mathrm{d}L_1/\mathrm{d}x_1$ (or $\mathrm{d}L_3/\mathrm{d}x_3$), respectively.
\begin{figure}
\begin{center}
    \includegraphics[width=7.5cm]{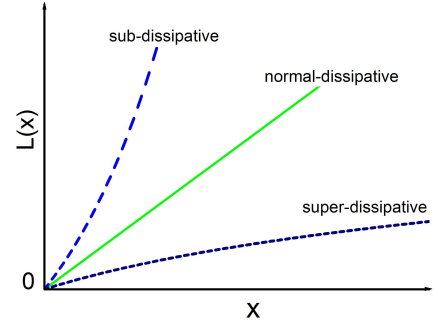}
  \caption{(Colored online) Schematic diagram of three types of dissipative engines.}\label{relation}
\end{center}
\end{figure}

\section{Optimization}
The heat absorbed or released by the engines can be expressed as
\begin{equation}\label{Q1}
    Q_1=T_1\Delta S_1- T_1L_1,
\end{equation}
and
\begin{equation}\label{Q3}
    Q_3=T_3\Delta S_1+ T_3L_3.
\end{equation}
Given the two equations above, we can obtain the efficiency
\begin{equation}\label{efficiency}
    \eta=1-\frac{Q_3}{Q_1}=\frac{\eta_C\Delta S_1-L_1-(1-\eta_C)L_3}{\Delta S_1-L_1},
\end{equation}
and the power
\begin{equation}
 P=\frac{[(T_1-T_3)\Delta S_1 -(T_1L_1+T_3L_3)]x_1x_3}{(x_1+x_3)}.
\label{powerr}
\end{equation}

Maximizing the power with respect to $x_1$ and $x_3$, we can obtain
 \begin{equation}\label{opt1}
    [(T_1-T_3)\Delta S_1 -(T_1L_1+T_3L_3)]x_3=T_1x_1(x_1+x_3)L'_1,
 \end{equation}
and
 \begin{equation}\label{opt2}
    [(T_1-T_3)\Delta S_1 -(T_1L_1+T_3L_3)]x_1=T_3x_3(x_1+x_3)L'_3.
 \end{equation}
Dividing Eqs.~\eqref{opt1} and \eqref{opt2}, we derive
 \begin{equation}\label{GCR}
    T_1x_1^2L'_1=T_3x_3^2L'_3.
 \end{equation}
Adding Eqs.~\eqref{opt1} and \eqref{opt2} with the consideration of Eq.~\eqref{efficiency}, we find that the EMP satisfies
\begin{equation}\label{EMP}
    \eta^\ast=\frac{\eta_C}{1+\frac{(1-\eta_C)(L_1+L_3)}{x_1L'_1+(1-\eta_C)x_3L'_3}},
\end{equation}
which is the key equation in our paper.

\section{Bounds of EMP for three types of dissipative heat engines}
In this section, we will discuss the bounds of EMP for three types of Carnot-like dissipative engines in terms of the characteristics of the relation between the minimum entropy production in each ``isothermal" processes and the time for completing those processes.
\subsection{Normal-dissipative engines}
The minimum entropy production in the ``isothermal" processes and the time for completing those processes of normal dissipative engines satisfy $ x_1L_1' = L_1$ and $x_3L_3'= L_3$ which imply $x_1L'_1+(1-\eta_C)x_3L'_3=L_1+(1-\eta_C)L_3$. Considering $0<\eta_C<1$, we derive $(1-\eta_C)(L_1+L_3)<L_1+(1-\eta_C)L_3<L_1+L_3$. Considering Eq.~\eqref{EMP}, we can derive the EMP of normal-dissipative engines to be bounded between $\eta_-\equiv \eta_C/2$ and $\eta_+ \equiv \eta_C/(2-\eta_C)$ which are the same as the bounds obtained by Esposito and his coworkers \cite{Esposito2010}. A major difference is that here we derive the bounds directly from Eq.~\eqref{EMP} without calculating the explicit expression of EMP. Particularly, if we consider Eqs.~\eqref{GCR} and \eqref{EMP} for the symmetric normal-dissipation case $L_1'=L_3'$, it is not difficult for us to recover the CA efficiency $\eta_{CA}=1-\sqrt{1-\eta_c}$ in this special case. In addition, we should point out that the time for competing the two isothermal processes satisfied the relation that $t_1/t_3=\sqrt{T_1/T_3}$ in this symmetric normal-dissipation case.
\subsection{Sub-dissipative engines}
The  sub-dissipative engines satisfy $x_1L_1'> L_1$ and $x_3L_3'> L_3$ which imply $x_1L'_1+(1-\eta_C)x_3L'_3>L_1+(1-\eta_C)L_3$. Given that $L_1+(1-\eta_C)L_3>(1-\eta_C)(L_1+L_3)$, then we finally derive the lower bound of EMP to be $\eta_{-}\equiv\eta_C/2$ for the sub-dissipative engines from Eq.~\eqref{EMP}. The above inequalities give no confinement on the upper bound, thus we may take $\eta_+\equiv\eta_C$ as a reasonable estimate.
\subsection{Super-dissipative engines}
The super-dissipative engines satisfy $x_1L'_1  < L_1$ and $ x_3L'_3  < L_3$ which imply $x_1L'_1+(1-\eta_C)x_3L'_3<L_1+(1-\eta_C)L_3$. Given that $L_1+(1-\eta_C)L_3<L_1+L_3$, then we finally derive the upper bound of EMP to be $\eta_{+}\equiv\eta_C/(2-\eta_C)$ for the super-dissipative engines. The above inequalities give no confinement on the lower bound, thus we may take $\eta_-\equiv 0$ as a conservative estimate.
\subsection{Examples for three types of engines}
For examples, we consider the relation of power-law profile, $L_i= \Gamma_i x_i^n,  (i=1,3)$ where $\Gamma_i>0$ and $n>0$ are given parameters. It is easy to see that a heat engine is of normal-, sub- or super-dissipative type if $n=1$, $n>1$ or $0<n<1$, respectively. Substituting this relation into Eqs.~\eqref{GCR} and \eqref{EMP}, we can explicitly derive the expression of EMP as
\begin{equation}\label{etan}
    \eta^\ast=\frac{\eta_C}{1+\frac{1}{n}-\frac{\eta_C}{n[1+(T_3\Gamma_3/T_1\Gamma_1)^{1/(n+1)}]}},
\end{equation}
from which we can find
 \begin{equation}\label{bound}
  \frac{\eta_C}{1+1/n}<\eta^\ast<\frac{\eta_C}{1+1/n-\eta_C/n}.
 \end{equation}
For normal-dissipative engines, $n=1$, Eq.~\eqref{bound} implies that $\eta_C/2<\eta^\ast<\eta_C/(2-\eta_C)$. For  sub-dissipative engines, $n>1$, Eq.~\eqref{bound} implies that $\eta_C/2 < n\eta_C/(n+1)<\eta^\ast<n\eta_C/(n+1-\eta_C)< \eta_C$ where the upper bound $\eta_C$ can be reached for sufficiently large $n$ while the lower bound can be reached for $n\rightarrow 1_+$. For super-dissipative engines, $0<n<1$, Eq.~\eqref{bound} implies that $0 <n\eta_C/(n+1)<\eta^\ast<n\eta_C/(n+1-\eta_C)<\eta_C/(2-\eta_C)$ where the lower bound $0$ can be reached for small enough $n$ while the upper bound can be reached for $n\rightarrow 1_-$.

The maximum power can also be calculated with consideration of the power-law profile, $L_i= \Gamma_i x_i^n$. From Eq.~\eqref{GCR} we can obtain that the inverse of time for completing the two processes satisfied $x_1/x_3=({\Gamma_3 T_3}/{\Gamma_1T_1} )^{\frac{1}{n+1}}$. Substituting this relation into Eq.~\eqref{opt2} to solve $x_1$ and $x_3$, then considering Eq.~\eqref{powerr}, we derive the maximum power output
{\small\begin{equation}\label{eq-maxpowern}
P_{\max}=\frac{n}{(n+1)^{1+1/n}}\left[\frac{(T_1-T_3)\Delta S_1}{(T_1\Gamma_1)^{1/(n+1)}+(T_3\Gamma_3)^{1/(n+1)}}\right]^{1+1/n}.
\end{equation}}
When $n=1$ and $\Gamma_3=\Gamma_1$, the above equation reduces to $P_{\max}=(\Delta S_1^2)/(4\Gamma_1)(\sqrt{T_1}-\sqrt{T_3})^2$ which is in agreement with the universal behavior [$\simeq (\sqrt{T_1}-\sqrt{T_3})^2$] of maximum power obtained by Curzon and Ahlborn~\cite{Curzon1975}.
On the other hand, for large $n$, equation \eqref{eq-maxpowern} degenerates into $P_{\max}\simeq(T_1-T_3)\Delta S_1/2(\Gamma_1/\Delta S_1)^{1/n}$, which implies that the maximum power of sub-dissipative engines is still nonvanishing although its EMP can approach to the Carnot efficiency.

\section{conclusion and discussion}
The heat engines are classified into three dissipative types according to characteristics of the relations between the minimum irreversible entropy production in the ``isothermal" processes and the time for completing those processes. The bounds of EMP for three types of dissipative Carnot-like engines
can be summarized as
{\small \begin{equation}\label{etamaxp}
    \begin {cases}
    \eta_C/2<\eta^\ast<\eta_C/(2-\eta_C), & \mathrm{normal-dissipative} \\
    \eta_C/2< \eta^\ast< \eta_C              , & \mathrm{sub-dissipative}\\
    0<\eta^\ast<\eta_C/(2-\eta_C)                , & \mathrm{super-dissipative}
    \end{cases}\nonumber
\end{equation}}
which display certain universality for each types of dissipative Carnot-like engines. It is interesting that the bounds of EMPs for normal-, sub- and super-dissipative Carnot-like engines correspond respectively to those for linear, sub- and super-linear irreversible Carnot-like engines discussed in our previous work \cite{TuW201110}. However, a simple consideration implies that these two kinds of classifications are different from each other \cite{TuW201108}. It is still a challenge to understand why they share the same bounds.

Another major challenge is the entropy production in the adiabatic process which is assumed to be vanishing in most of researches~\cite{Salamon80,Salamon80,Chen1989,ChenJC94,Bejan96,ChenL99,vdbrk2005,dcisbj2007,Schmiedl2008,Tu2008,Chenlin2009,
Esposito2009a,Esposito2009,Esposito2010,EspositoPRE10,GaveauPRL10,Izumida,wangx10,Velasco10,Abe2011,chenl2011,Izumida2012,
Schmiedl07,TuW201108,WangHe,WangHe41148,Apertet}. This is indeed true for some micro-systems such as the stochastic heat engines proposed by Schmiedl and Seifert \cite{Schmiedl2008}. However, it might not be true for the macroscopic heat engines and quantum heat engines. Thus the time for complete two adiabatic transitions should not be neglected. Recently, Wang and He argued that the entropy production might be also inversely proportional to the time for completing the adiabatic process \cite{wangheinter}. It is necessary for us to investigate this issue carefully in the further work.

\section{acknowledgement}
The authors are grateful to the financial support from Nature Science Foundation of China (Grant NO.11075015).

\end{document}